\documentclass{article}

\usepackage[english]{babel}

\usepackage[letterpaper,top=2cm,bottom=2cm,left=3cm,right=3cm,marginparwidth=1.75cm]{geometry}

\usepackage{amsmath}
\usepackage{graphicx}
\usepackage[colorlinks=true, allcolors=blue]{hyperref}
\usepackage{caption}
\usepackage{subcaption}

\title{Design of a plasma chamber for a high-density 10 kW ECR ion source}
\author{I. Izotov$^1$, V. Skalyga$^1$, A. Bokhanov$^1$\\ \\
$^1$Federal Research Center Institute of Applied Physics\\
of the Russian Academy of Sciences\\
46 Ul'yanov Street, Nizhny Novgorod, 603950, Russia}
\date{}

\begin{document}
\maketitle

\begin{abstract}
Gasdynamic electron cyclotron resonance ion sources are known for their ability to produce huge currents of low to moderately charged ion beams with superior quality. This is achieved by means of tens of kW of microwave heating power, leading to high plasma density. Average power density in the plasma chamber may reach hundreds of W/cm$^3$, which poses very high thermal load to plasma chamber walls. A water-cooled plasma chamber has been developed for the GISMO gasdynamic ECRIS, which is able to handle 10 kW of input microwave power with average power density in the plasma of 250 W/cm$^3$. The plasma chamber was manufactured and successfully tested for 200 hours.
\end{abstract}

Gasdynamic ECR ion sources were developed at the Institute of Applied Physics of the Russian Academy of Sciences (IAPRAS) within studies on an ECR discharge in a simple mirror magnetic trap heated by powerful gyrotron radiation in the frequency range of 24-75 GHz. These investigations were motivated by an attempt to improve a performance of the conventional source of multiply charged ions (also known as ECRIS) by significant increase of the heating radiation frequency in accordance with the well-known scaling laws \cite{Geller}. Gasdynamic sources utilize a high plasma density and use a different plasma confinement mechanism when compared to the conventional ones. Due to high collision frequencies, this confinement mode is governed mainly by a quasi-gasdynamic (hereinafter referred to as gasdynamic) plasma outflow through the magnetic trap mirrors, opposite to the collision-less confinement of a rarefied plasma in conventional ECRISes. The lifetime of a gasdynamic plasma in a compact magnetic trap is on the order of several tens of microseconds, which is hundreds of times shorter than in a collisional-less ECRIS. High losses dictate that an average electron energy usually is on the level of 50-100 eV even at 100 kW of heating power, whereas conventional ECRIS demonstrate 1-10 keV at several kW. Pointless for a high charged ions production, plasma with density in the range of 10$^{13}$ - 10$^{14}$ cm$^{-3}$ and the electron temperature 10-50 eV turned out to be extremely advantageous for the light ion beams production: the gasdynamic source demonstrated beams of light ions with very high currents and exceptional quality, i.e. record-breaking brightness \cite{GECRIS}.

\begin{figure}
	\centering
	\includegraphics[width=0.8\textwidth]{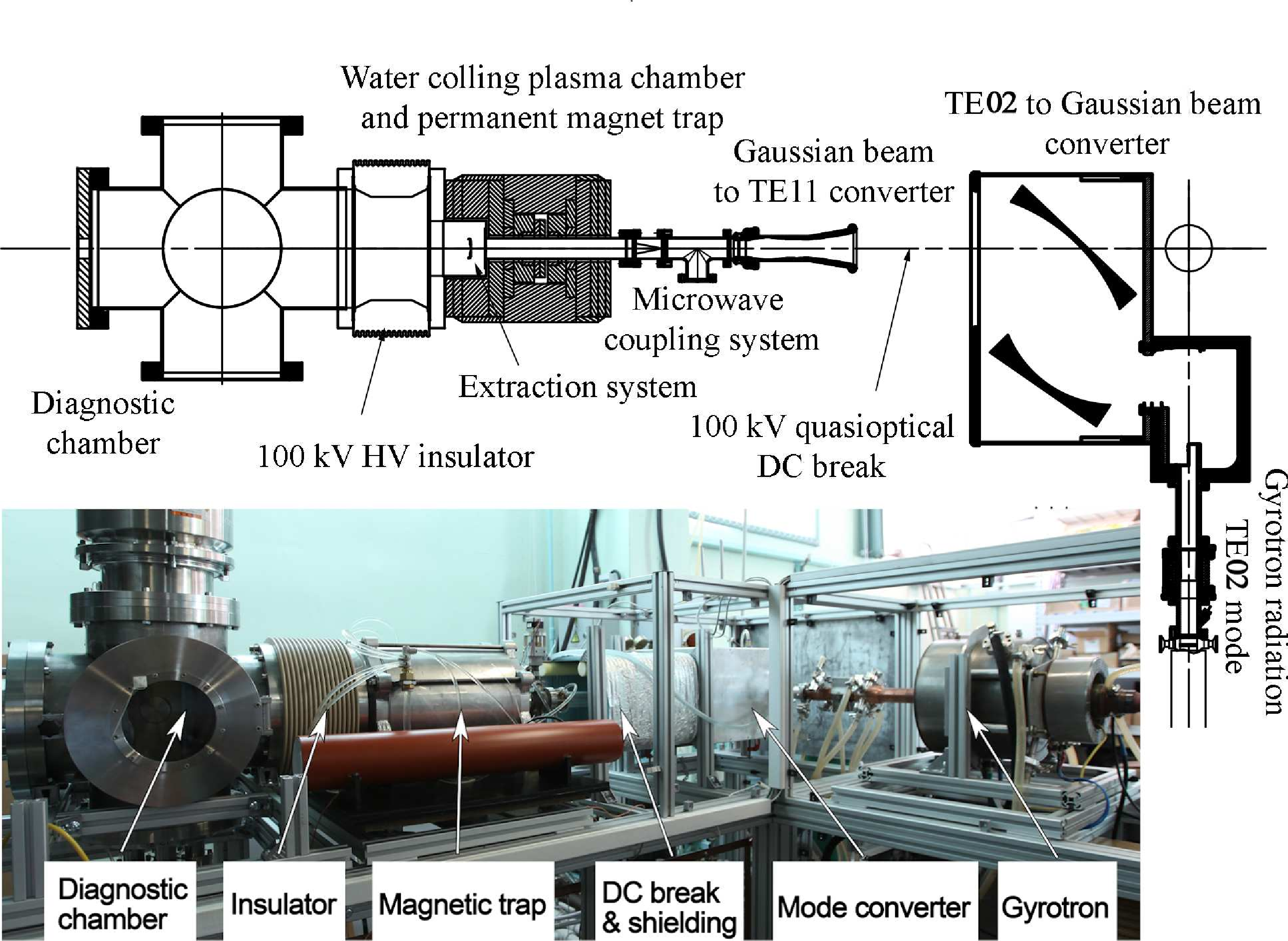}
	\caption{\label{fig1}Overview of the GISMO experimental facility.}
\end{figure}

These results stimulated the construction of the CW gasdynamic ion source named GISMO \cite{GISMO}. The general scheme of the facility is shown in Fig.~\ref{fig1}. A 28 GHz gyrotron is used for the plasma heating with maximum power of 10 kW. The plasma is confined within a simple mirror-like magnetic trap built with permanent magnets. The magnetic field strength is 1.5 T at mirrors and 0.25 T at the trap center, yielding the mirror ratio of 6. The distance between magnetic mirrors is 120 mm. The magnet bore is 50 mm. An ion beam extraction system was designed to be able to sustain upt to 100 kV of accelerating voltage. Instead of a conventional waveguide DC break, a quasi-optical system is used due to a very high potential difference. The electromagnetic radiation of the gyrotron (TE$_{02}$ mode of a circular waveguide) is converted to the gaussian beam at ground potential, delivered over air to the horn antenna and converted into a TE$_{11}$ mode of a circular waveguide at high potential. The main goal of studies at GISMO facility is the development of a powerful D-D neutron generator \cite{DD} capable of producing the neutron flux sufficient for boron neutron capture therapy (BNCT), which implies the ability to produce a hundreds of mA of deuterium beam with superior quality (i.e. low emittance), enabling the ability to deliver and focus the beam onto the target. 

One of distinctive features of GISMO facility is a high power density. The plasma volume is estimated to be on the order of 50 cm$^3$, which yields up to 250 W/cm$^3$ at the maximum available heating power of 10 kW, whereas conventional ECRISes are usually operate at 1-5 W/cm$^3$ \cite{HIISI}. As was mentioned earlier, such level of power allows to run the ion source in gasdynamic mode, while sustaining the electron energy high enough for close to 100\% ionization. Possibility of that is confirmed by, for example, results obtained with helium beam, where tens of mA of He$^{2+}$ was demonstrated, and the average charge was more than 1.5 \cite{Skalyga_review}.
 
However, present design of the plasma chamber doesn't allow to use all of the available power in continuous mode of operation on a long timescale: the plasma collectors overheat at the power above 5 kW, if the operating time is more than several minutes. This leads to a need of redesigning the plasma chamber and especially, its cooling circuits. The problem is further complicated by the fact that the vacuum bore of the plasma chamber cannot be smaller than 32 mm due to the waveguide breakdown electric field magnitude, and the maximum outer diameter must be below 50 mm (magnet bore).

\begin{figure}
	\centering
	\includegraphics[width=0.8\textwidth]{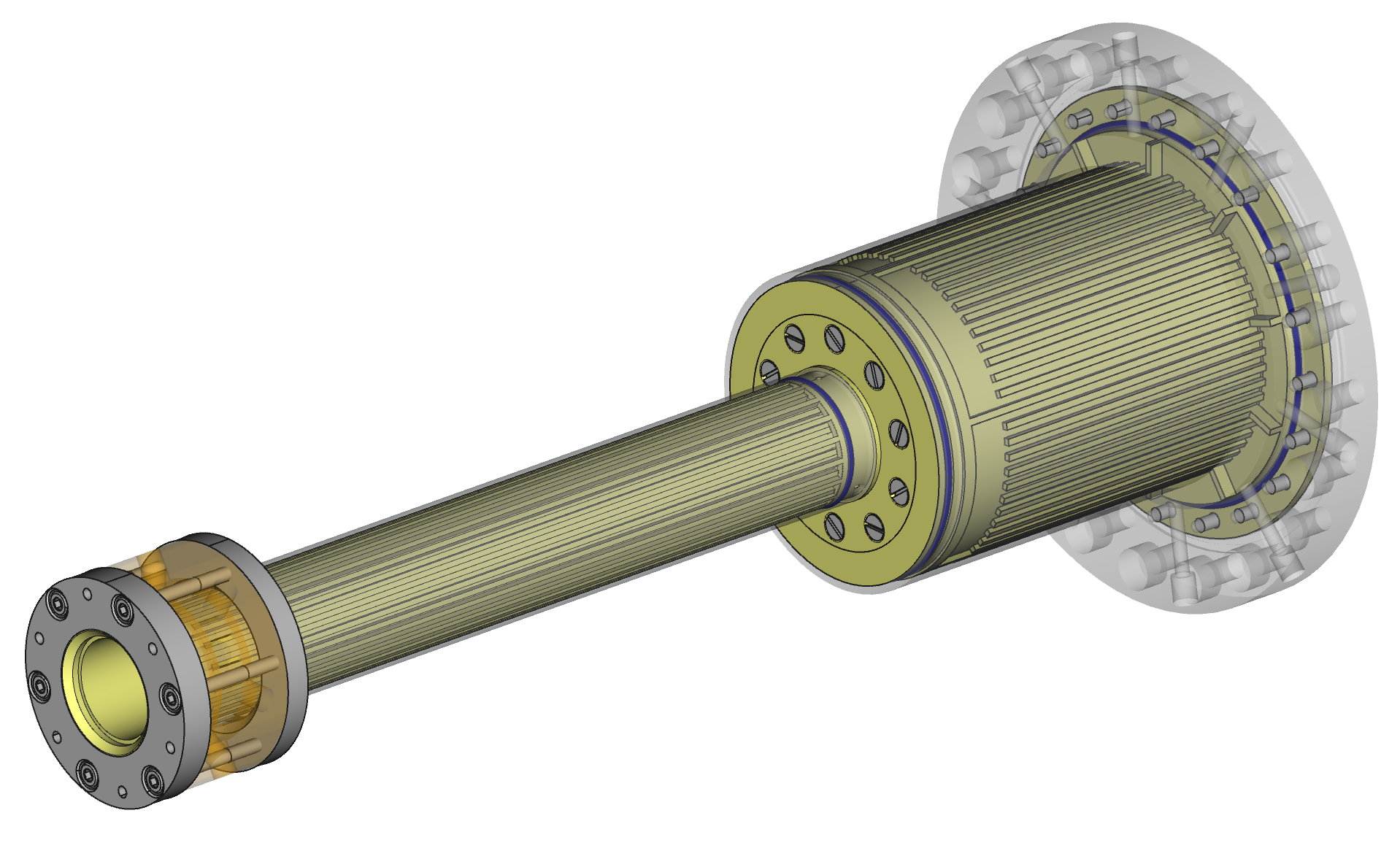}
	\caption{\label{fig2}3d rendering of the developed plasma chamber.}
\end{figure}

Solution was found in the use of a layered walls with a micro-channel-like cooling, and a separate microwave launching system, which acts as a plasma collector. A general view of the new plasma chamber is shown in Figure~\ref{fig2}. For clarity, the cooling system case is made translucent. The main part of the plasma chamber is a cylinder with a bore of 36 mm. Next, it is sealed to the so-called expander, which fits into corresponding opening in the magnetic system, designed to accommodate extraction system electrodes and various diagnostics. Both plasma and diagnostic parts are cooled using longitudinal grooves in the chamber walls 3 mm deep and 2 mm wide. The walls in contact with the plasma are made of brass. On the one hand, brass has a significantly lower thermal conductivity (depending on the exact composition, in the range of 100 - 200 W / (m * K)) than pure copper, however, calculations have shown that even in the case of minimal thermal conductivity, the proposed system, which has a developed cooled surface, provides a safe temperature in all key nodes. At the same time, brass has a fundamentally greater mechanical strength and rigidity, which makes it possible to make the minimum wall thickness at the vacuum-water or vacuum-atmosphere boundary 1.5 mm. The total thickness of the brass walls (excluding grooves) is 5 mm. On top of the brass cylinders, external stainless steel cases with a wall thickness of about 1.5 mm are put on, which simultaneously act as a stiffening element and a wall of the cooling system. The atmosphere-water interface was sealed using radial sealing rings. The cooling system is divided into two circuits: the first circuit is cooling the plasma chamber, water inlet and outlet are situated in the frontal flow-separating module, which is also radially sealed (see the left in Fig. 2). The second circuit similarly cools the expander chamber, the water inlet and outlet is organized in the thickness of the CF-160 flange, which connects the plasma chamber to the rest of the vacuum volume.

\begin{figure}
	\centering
	\includegraphics[width=0.8\textwidth]{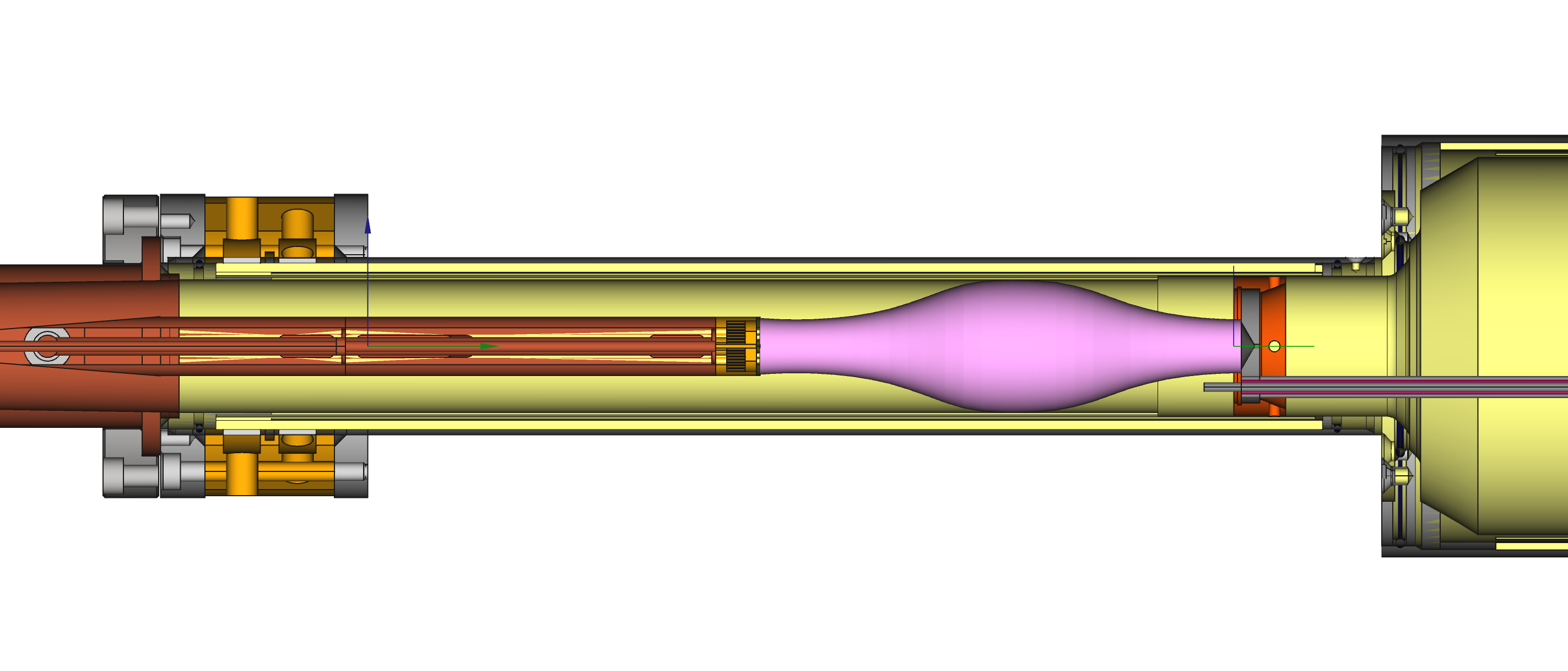}
	\caption{\label{fig3}Cut view of the plasma chamber with injection and extraction plugs installed. Pink spindle demonstrates assumed plasma volume.}
\end{figure}

The magnetic field confines the plasma in radial direction while allowing a free flow axially. Thus, the highest thermal load appears in axial direction as well. At GISMO, one of the axial walls acts as plasma collector, part of the microwave launching system and gas injection point simultaneously, and referred to as ``injection plug''. Accordingly, the ``extraction plug'' is the axial wall, or plasma collector, placed at the opposite side of the magnetic trap, and acts as a plasma electrode - a part of the ion beam formation system. The ``extraction plug'' is much simpler than its counterpart, as its only function is to provide a conductive wall for the plasma, having a small aperture for the ion beam formation. This is demonstrated in Figure~\ref{fig3}, where the pink shape is a visualization of a magnetic lines that touch the walls and confines the bulk of the plasma.

\begin{figure}
	\centering
	\begin{subfigure}[b]{0.3\textwidth}
		\centering
		\includegraphics[width=\textwidth]{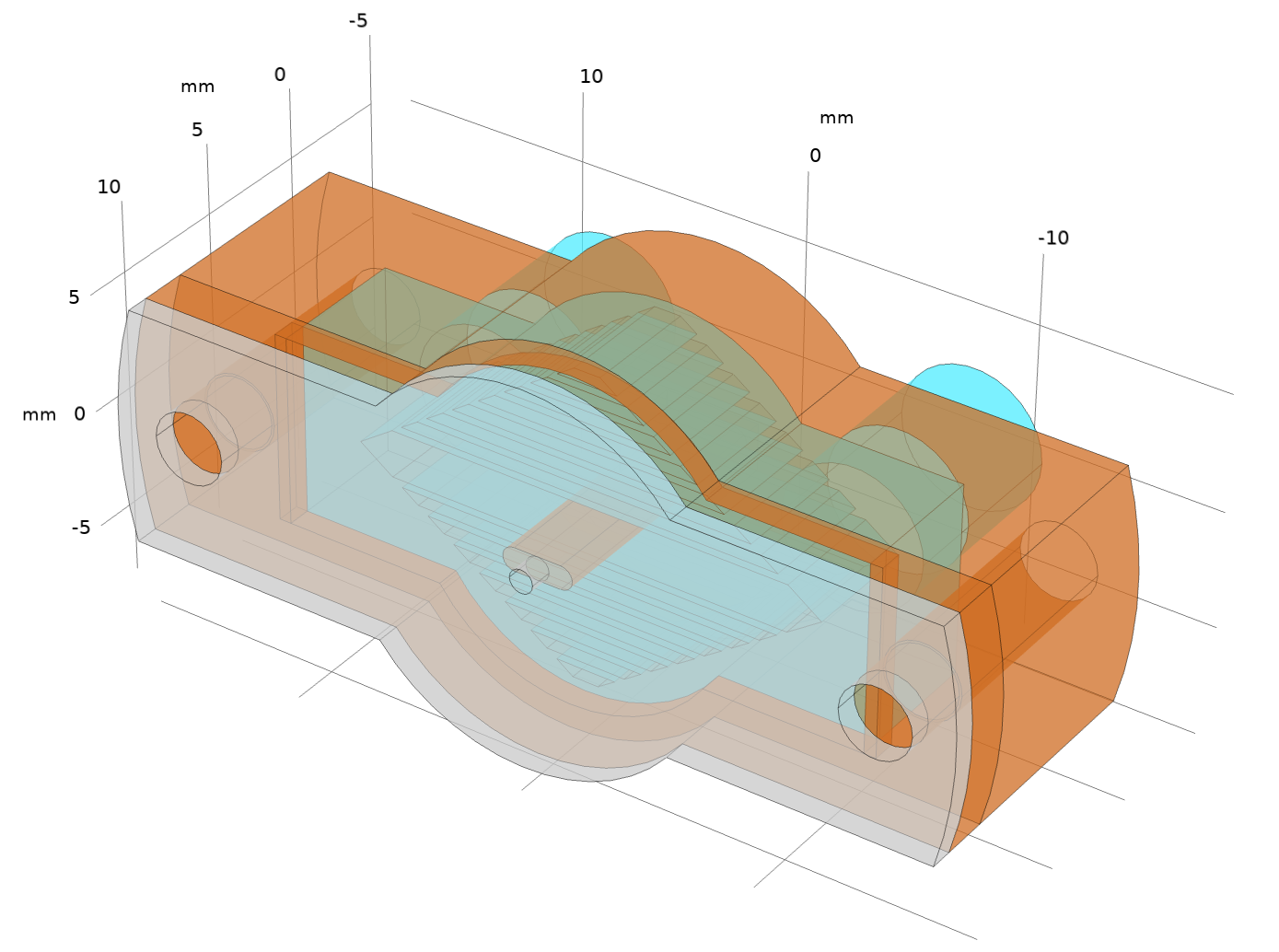}
		\caption{}
		\label{fig4a}
	\end{subfigure}
	\hfill
	\begin{subfigure}[b]{0.3\textwidth}
		\centering
		\includegraphics[width=\textwidth]{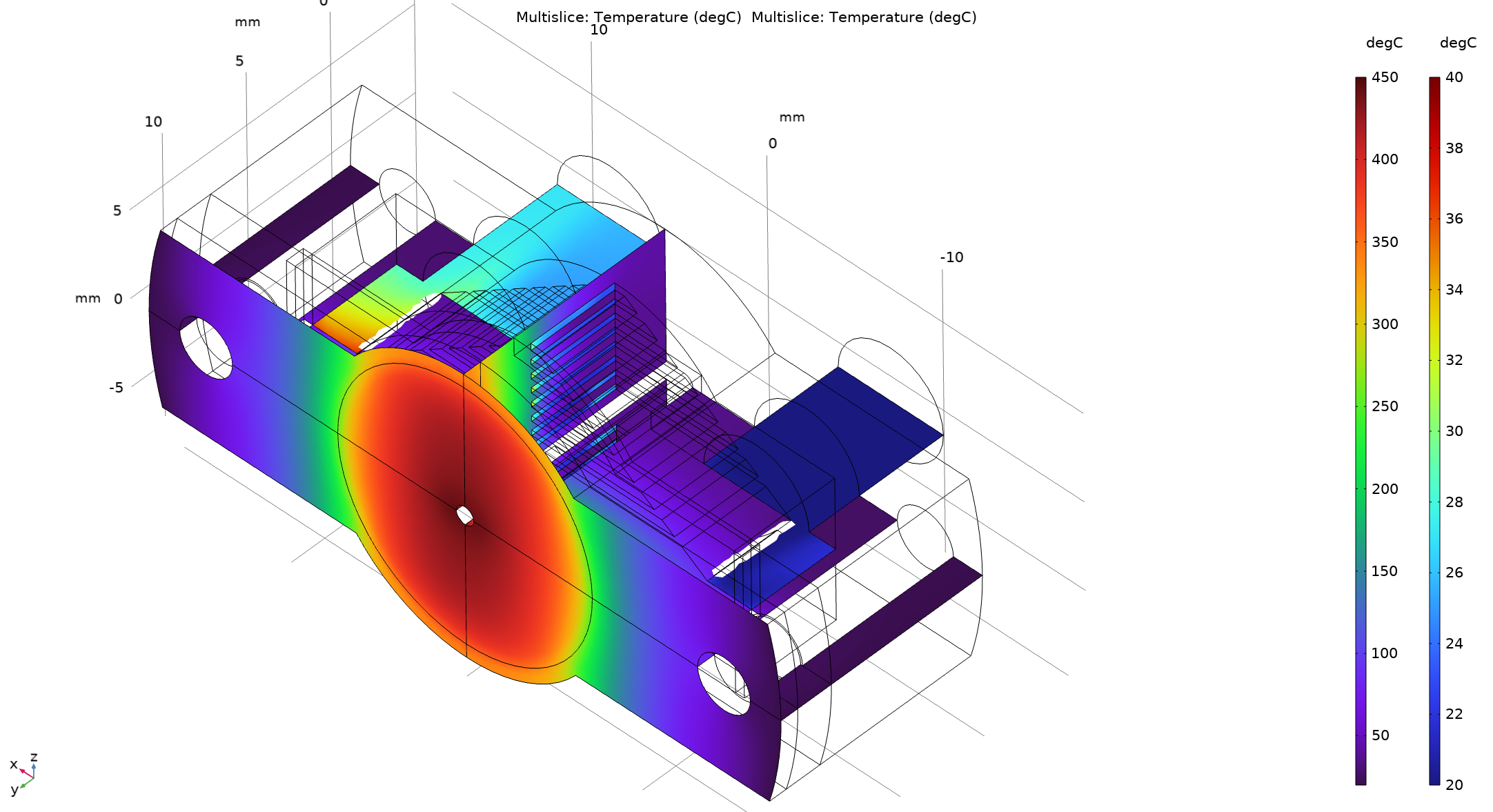}
		\caption{}
		\label{fig4b}
	\end{subfigure}
	\hfill
	\begin{subfigure}[b]{0.3\textwidth}
		\centering
		\includegraphics[width=\textwidth]{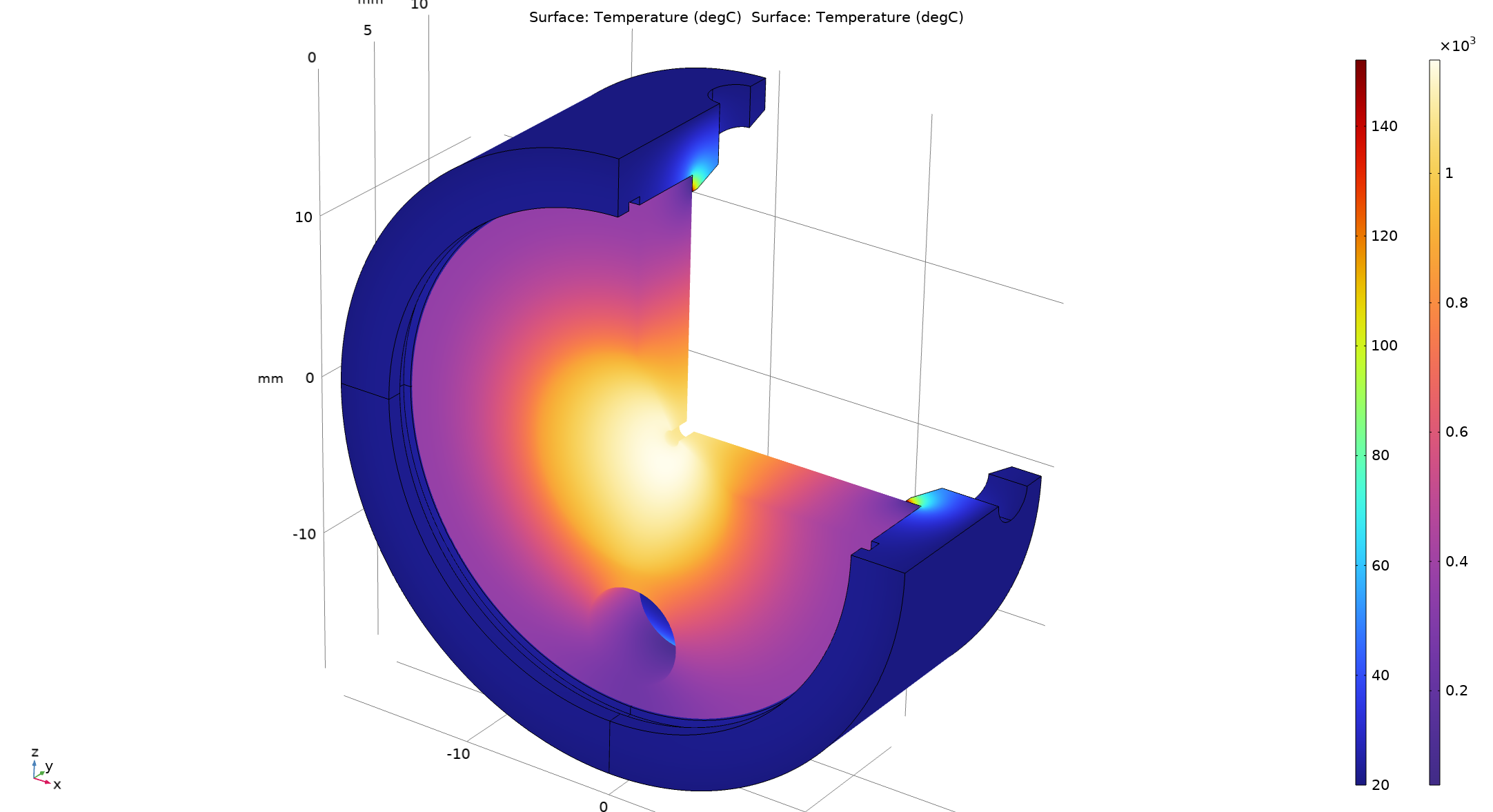}
		\caption{}
		\label{fig4c}
	\end{subfigure}
	\caption{(a) Structure of the injection plug end. Blue: water domain; orange: copper domain, gray: molybdenium domain.
	(b) Temperature distribution in the injection plug end. (c) Temperature distribution in the extraction plug end.}
	\label{fig4}
\end{figure}

As a result of numerical optimization, the most optimal geometry of the injection plug end was found, which made it possible to achieve the maximum heating power while maintaining a safe temperature level. The geometry is shown in Figure~\ref{fig4a}. The copper part is shown in orange, molybdenum in gray, water in blue, i.e. cooling channels inside the walls. The geometry of the outer perimeter is dictated by electrodynamics, i.e. reflection-less transmission of microwave radiation. The calculation of the heat load was carried out by jointly solving the equations of heat conduction and hydrodynamics on a tetrahedral mesh using the finite element method under the following boundary conditions: inlet water temperature - 20 degrees, inlet water pressure - 4 bar, heat flow to the face - 5 kW, all walls are thermally insulated. The last boundary condition is too strong, since part of the walls has thermal contact with the walls of the plasma chamber and, accordingly, is also cooled, but at the same time it certainly provides an upper estimate for the stationary temperature.

\begin{figure}
	\centering
	\includegraphics[width=0.8\textwidth]{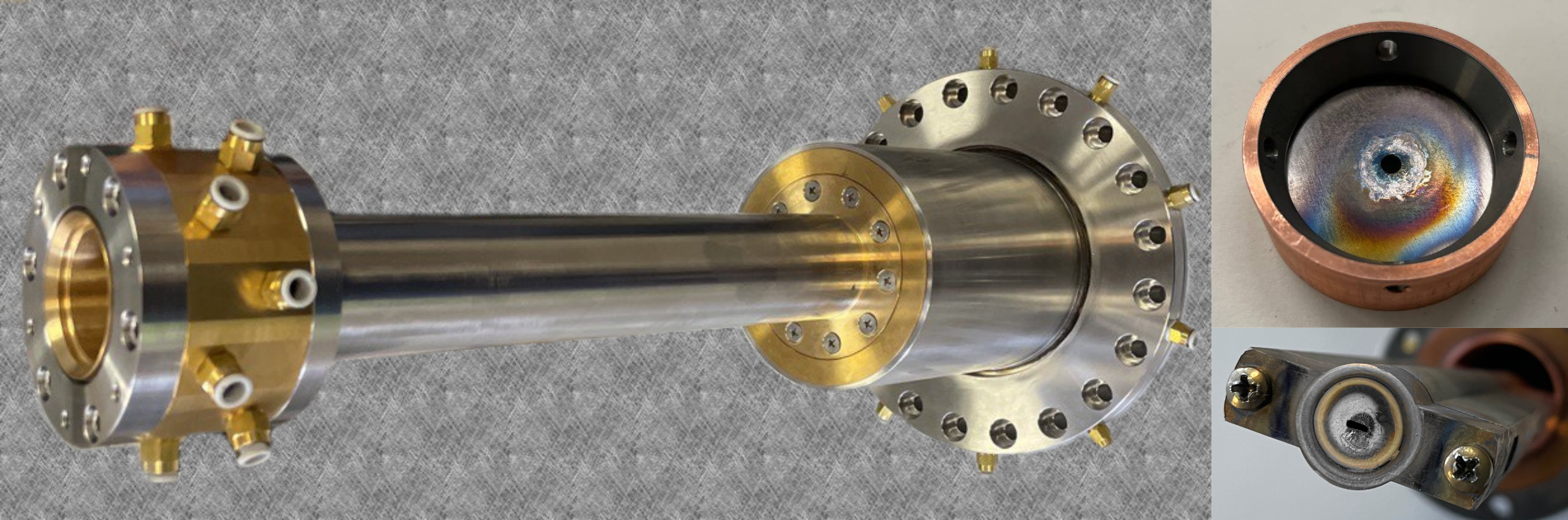}
	\caption{\label{fig5}(a) manufactured and assembled plasma chamber. Extraction (b) and injection (c) plugs after 200 hours of 10 kW operation.}
\end{figure}

Figure~\ref{fig4b} shows the stationary temperature distribution in the water domain and on the surface of plasma collector. Due to the fundamentally different scale, two color scales are used in the figure: the left one corresponds to the distribution on the molybdenum surface on which the plasma flows, and the right one corresponds to the water temperature. It can be seen that the water temperature never exceeds 40 \textdegree C, and the maximum temperature of the molybdenum surface is 450 \textdegree C, which is a safe value for both molybdenum and copper radiator, touching the molybdenum plate.

The extraction plug is much simpler technologically, since it is, in fact, a simple diaphragm. A passive method was used to cool the diaphragm: it is made of molybdenum 5 mm thick and pressed into a copper holder, which is inserted into the plasma chamber. When heated under the influence of the plasma flow, thermal expansion of the diaphragm and holder occurs, which ensures reliable thermal contact with the cooled wall of the plasma chamber. The calculation of the stationary distribution of the temperature of the diaphragm and the wall of the plasma chamber was carried out in the same way as for the injection plug. Figure~\ref{fig4c} shows the temperature distribution for the same heat load of 5 kW. It is seen that the hottest point has the temperature of 1200 \textdegree C and is situated, quite obviously, close to the diaphragm center. The maximum temperature of the copper holder never reaches above 150 \textdegree C.

The designed system was manufactured at IAP RAS. Figure~\ref{fig5}a shows a photograph of the assembled plasma chamber. A number of test experiments were carried out with a discharge in nitrogen and argon, aimed at testing new elements under a significant thermal load. The tests were carried out for several weeks at about 6 hours a day. At the beginning of each test day, a discharge was ignited at a low power (about 2 kW), after which the power was gradually increased to 10 kW, while the parameters of the cooling system were controlled. The total operating time during the tests reached 200 hours. Figure~\ref{fig5}b and \ref{fig5}c shows photographs of plasma collectors after thermal testing has been completed. There are traces of slight erosion on the collectors, but its magnitude is apparently no more than a few microns. Thus, the developed system is deemed to be ready for a long-term operation at 10 kW of input microwave power.

\section*{Acknowledgments}
The work was supported by the Russian Science Foundation, project 19-12-00377.

\bibliographystyle{ieeetr}
\bibliography{main}

\end{document}